\documentclass[aip,cha,reprint,floatfix]{revtex4-2}

\usepackage{amsmath,amssymb}
\usepackage{color}
\usepackage{fancybox}
\usepackage{graphicx}

\usepackage[utf8]{inputenc}
\usepackage[T1]{fontenc}
\usepackage[english]{babel}
\usepackage{eurosym}

\usepackage[colorlinks]{hyperref}

\usepackage{enumerate}
\usepackage{soul,array}
\usepackage{pifont}

\usepackage{multirow}

\usepackage{fancyvrb}

\begin{document}

\title{
Recurrent patterns of user behavior in different electoral campaigns: A Twitter analysis of the Spanish general elections of 2015 and 2016
}
\date{\today}
\keywords{twitter; user behavior; temporal networks; complex networks; elections; spain}
\author{S. Martin-Gutierrez}
\affiliation{Grupo de Sistemas Complejos,
Escuela T\'ecnica Superior de Ingenier\'ia Agron\'omica, 
Alimentaria y de Biosistemas,
Universidad Polit\'ecnica de Madrid,
Avda.~Complutense s/n 28040 Madrid, Spain.}
\author{J.C. Losada}
\affiliation{Grupo de Sistemas Complejos,
Escuela T\'ecnica Superior de Ingenier\'ia Agron\'omica, 
Alimentaria y de Biosistemas,
Universidad Polit\'ecnica de Madrid,
Avda.~Complutense s/n 28040 Madrid, Spain.}
\author{R. M. Benito}
\affiliation{Grupo de Sistemas Complejos,
Escuela T\'ecnica Superior de Ingenier\'ia Agron\'omica, 
Alimentaria y de Biosistemas,
Universidad Polit\'ecnica de Madrid,
Avda.~Complutense s/n 28040 Madrid, Spain.}

\begin{abstract}
We have retrieved and analyzed several millions of Twitter messages corresponding to the Spanish General elections held on the 20th of December 2015 and repeated on the 26th of June 2016. The availability of data from two electoral campaigns that are very close in time allows us to compare collective behaviors of two analogous social systems with a similar context. By computing and analyzing the time series of daily activity, we have found a significant linear correlation between both elections. Additionally, we have revealed that the daily number of tweets, retweets and mentions follow a power law with respect to the number of unique users that take part in the conversation. Furthermore, we have verified that the topologies of the networks of mentions and retweets do not change from one election to the other, indicating that their underlying dynamics are robust in the face of a change in social context. Hence, in the light of our results, there are several recurrent collective behavioral patterns that exhibit similar and consistent properties in different electoral campaigns.
\end{abstract}

\maketitle

\section{Introduction}

Nowadays, social networking sites (SNS) are a well established communication medium. They are used by a huge user base to share experiences, discuss opinions, read the news, etc. Twitter is one of the most dynamic SNS with respect to the interactions among users and one of the most powerful with respect to the potential information that can be extracted for research purposes. Moreover, in the last quarter of 2015 this social network had 305 million monthly active users, and the last data show that the number of monthly active users has risen to 335 million in the second quarter of 2018 \cite{statista_twitter_users}.

These facts have stimulated the developement of research projects from a wide variety of fields, from sociology to network science,  to study user behavior under new perspectives. These studies unravel the emergent patterns in our collective behavior and show how they can be used to gain insight about relevant social topics like economy \cite{10.1371/journal.pone.0128692}, marketing \cite{asur:2010:pfs:1913793.1914092}, politics \cite{ICWSM101441, doi:10.1063/1.4729139},  mobility \cite{doi:10.1080/15230406.2014.890072} or polarization \cite{ICWSM112847, doi:10.1063/1.4913758}.

In this work, we are interested in electoral processes. In these kind of contexts, people use social media as a communication medium to exchange opinions. However, there are also some users whose purpose is to influence the conversation in a way that may affect the voting choices of the people. Although these are online actions, they have an impact on the offline world.

The concise character of a tweet turns it into a powerful tool to share breaking news and take part in dynamic debates \cite{CPLX:CPLX21457}. That is why it is one of the most used media by journalists, politicians and politics enthusiasts to share their views and news. Nowadays, even traditional media often use Twitter as a news source. Furthermore, politicians consider Twitter one of the most consequential social media for their job \cite{doi:10.1386/jmpr.14.4.291_1}, although other online networking sites like Facebook also hold a high relevance.

In the Obama campaign of 2008, the efficacy of Twitter as a communication medium in a political context was established for the first time\cite{doi:10.1080/10810730903033000, PSQ:PSQ3815}. From then on, it has gained more and more relevance in political campaigns. This trend has also inspired an ever growing number of academic works that use Twitter data to analyze different aspects of the political campaigns.

Some lines of research are centered around predicting the outcome of elections following diverse techniques \cite{ICWSM101441, doi:10.1063/1.4729139, 10.1371/journal.pone.0095809}. Other works investigate the interaction among parties to assess the cohesion of the parties \cite{10.1371/journal.pone.0166586} or the level of debate \cite{doi:10.1063/1.4729139}. There are also studies that analyze the content of the tweets in order to detect their sentiment \cite{Yaqub2017, 7752220}. Two extensive reviews that cover a variety of subjects within the area can be found in \cite{doi:10.1177/0894439313493979, doi:10.1080/19331681.2015.1132401}.

In a previous paper \cite{doi:10.1063/1.4729139} the Spanish general elections of 2011 were studied using Twitter data. That work was focused on the prediction of the outcome of the election using a metric that measures political sentiment. The authors also explore the interactions among users by analyzing the structural and dynamical patterns of the complex networks emergent from the mention and retweet networks. The communication dynamics among politicians are also studied, finding a lack of debate, and a network growth model is proposed to reproduce these interactions. 

Although we compare some of our results to those presented in that work, in this study we do not aim to predict the outcome of the elections or simply study the relations among users and politicians. We take advantage of the availability of Twitter data gathered during two electoral campaigns (the Spanish general elections of 2015 and the repetition of the elections in 2016) that are very close in time to compare the collective user behavior manifested in two analogous social systems with a similar context. The individual activity and the political actors may be different; in fact, in the second election two parties formed a coalition, altering the political landscape. Our objective is to study and characterize emergent behaviors that are recurrently manifested in political contexts.

To this end, we have computed and compared time series of daily activity for both elections, finding that the temporal series of activity for both electoral periods are highly correlated. Furthermore, our results suggest that the number of tweets, retweets and mentions follow a power law with respect to the number of unique users that take part in the conversation (with an exponent slightly higher that 1). In order to explore the  evolution of the interactions among users, we have built networks of mentions and retweets and studied their temporal evolution. This has enabled us to verify that they show similar topological properties in both electoral periods. Besides, we have studied the mention and retweet subgraphs induced by political users and obtained results that imply a lack of communication among different parties and are in agreement with previous works. Furthermore, we have found that an intensification of the interaction can be detected between parties after the formation of a coalition.

The paper is organised as follows: In the second section we describe the political context, some relevant aspects of the interaction mechanisms in Twitter, the characteristics of our dataset and the methodology followed to build the networks of interactions. In the third section, we present and discuss our results with respect to the user activity and the evolution of the mention and retweet networks. We also use a metric to discuss the influence of regular users and politicians. Furthermore, we analyze the degree of debate among politicians of different parties. Finally, we summarize the main conclusions. 

\section{Materials and methods}

\subsection{Political context}
\label{sec_context}

Spain has a bicameral parliamentary system, where the lower house is called Congress of Deputies and the upper house, the Senate. For elections to the Congress of Deputies, held every four years, each of the 50 provinces serves as an electoral district, with the number of deputies representing it determined by its population. Under a proportional representation electoral system governed by the d’Hondt formula, ballots are cast for a provincewide party list rather than for candidates representing individual constituencies.

About four-fifths of the members of the Senate are directly elected via a plurality system at the provincial level. Each province is entitled to four representatives; voters cast ballots for three candidates, and those with the most votes are elected. The remainder of the senators are appointed by the regional legislatures. Because representation is not based upon population, in the Senate smaller and more-rural provinces generally are overrepresented in relation to their overall population \cite{britannica_elections}. Hence, with the end of illustrating the Spanish political landscape, in this work we have chosen to report only the results for the Congress of Deputies, as the distribution of seats is supposed to be more representative.

On the 20th of December 2015, the Spanish general elections were held. The PP (Partido Popular - People's Party) and the  PSOE (Partido Socialista Obrero Espa\~nol - Spanish Socialist Workers' Party), which constituted the traditional two-party system had lost a lot of social support, while the emerging parties Podemos (We can) and Cs (Ciudadanos - Citizens) were on the rise. This caused a transition from a two-party system to a multi-party system \cite{doi:10.1080/13608746.2016.1198454}.

In spite of that, the PP, which was holding the government, was still leading the polls. The rest of the parties were behind but not too far away. In fact, the supports for the other three main parties fluctuated so much during the year before the election that it seemed impossible to predict, by looking at the polls, which would be the final ranking of votes \cite{wiki_polls15}.

The result was a fragmented parliament where no party held an absolute majority and large coalitions were needed to form a government. The votes and seats that each party obtained are displayed in table \ref{tab_res_elec}. After several months of negotiations there was no agreement between any group of parties large enough to obtain a parliamentary majority that would allow the formation of a government. This led to the announcement of a new election on the 26th of June 2016.

\begin{table}[htbp]
\caption{Results of the Spanish general elections of the 20th of December 2015 and the 26th of June 2016. Podemos and IU are together in 2016 because they formed a coalition called UP.}
\begin{tabular}{|l|c|c|c|c|}
\hline
 & \multicolumn{ 2}{c|}{Votes} & \multicolumn{ 2}{c|}{Seats} \\ \cline{2-5}
 & {\bf 2015} & {\bf 2016} & {\bf 2015} & {\bf 2016} \\ \hline
PP & 7236965 & 7941236 & 123 & 137 \\ \hline
PSOE & 5545315 & 5443846 & 90 & 85 \\ \hline
Podemos & 5212711 & \multirow{2}{*}{5087538} & 69 &\multirow{2}{*}{71} \\ \cline{ 1- 2}\cline{ 4- 4}
IU & 926783 & \multicolumn{ 1}{c|}{} & 2 & \multicolumn{ 1}{c|}{} \\ \hline
Cs & 3514528 & 3141570 & 40 & 32 \\ \hline
Others & 2775011 & 2665069&26&25 \\ \hline
\end{tabular}
\label{tab_res_elec}
\end{table}

Before this new election, one of the emerging parties, Podemos, formed a coalition with IU (Izquierda Unida - United Left). This alliance was called UP (Unidos Podemos - United We Can). 

The 2016 election resulted in a parliament that was almost as fragmented as the one in 2015. The votes and seats obtained by each party are presented in table \ref{tab_res_elec}. In October 2016, the PP finally won the appointment vote and formed a minority government \footnote{In June 2018, a motion of no-confidence was issued against the ruling party (PP) and PSOE formed a minority government.}.

\subsection{Description of the data}
\label{sec_data}

We have worked with Twitter messages retrieved with the Twitter Streaming API. This API allows to download {\it tweets} matching a set of keywords. In order to avoid biased results, we have chosen the following neutral keywords to filter the messages: 

\begin{itemize}
\item Keywords for the 2015 election: {\it 20D, 20D2015, \#EleccionesGenerales2015}. 
\item Keywords for the 2016 election: {\it 26J, 26J2016, \#EleccionesGenerales2016, \#Elecciones26J}. 
\end{itemize}

We have downloaded tweets during a period of more than two months before and after each election. However, the core of our analysis has been focused on the 15 days of the official electoral campaign, the {\it reflection day} (day before the election), the election day and the day after. During that period of 18 days we have retrieved 1793145 {\it tweets} for the 2015 election and 1755438 for the 2016 election. 

Besides the keywords used to retrieve the data, the most used hashtags that appear in our dataset are the following:

\begin{itemize}
\item  Top hashtags of 2015:
\#podemos, \#psoe, \#partidopopular, \#ciudadanos, \#l6elecciones, \#7deldebatedecisivo, \#votapsoe, \#possible, \#pp, \#mivotocuenta, \#españa, \#españaenserio, \#podemos20dic, \#20dicpodemos, \#podemosremontada, \#votapp, \#hevotado, \#votapodemos20d.
\item Top hashtags of 2016:
\#afavor, \#unidospodemos, \#l6elecciones, \#votapsoe, \#Debate13j, \#psoe, \#cambioamejor, \#votapp, \#españa, \#ciudadanos, \#partidopopular, \#elcanvipossible, \#unsiporelcambio, \#avotar, \#lasonrisadeunpais, \#brexit, \#eleccionesgenerales.
\end{itemize}

We have also compiled lists of Twitter accounts associated to the four main parties (PP, PSOE, Podemos and Cs) and to IU. The latter is relevant because in the 2016 election, as we explained in section \ref{sec_context}, this party formed a coalition (UP) with Podemos, fact that is reflected on the data. These lists of accounts allow us to analyze the difference of behaviors of regular users and politicians. 

In order to build the set of political accounts, we have looked into the Twitter lists (which are lists of accounts elaborated by the users) defined by relevant official accounts associated to each party. We have downloaded those lists that include politicians, political institutions or supporters of the party. The total number of retrieved political users that participated in the conversation was 5227 in 2015, with an average number of followers of 4044 and 5012 in 2016, with an average number of followers of 4662.

\subsection{How does Twitter work}

In this section we will briefly describe some of the characteristics of Twitter. Specifically, the means of communication among users, which will be the source used to build the networks of interactions.

In Twitter there are several mechanisms of interaction among users. The first one is to {\it follow} other users. A user will receive every {\it tweet} that her followees post. Besides, there are five other mechanisms of interaction within the social network: the direct message, the {\it mention}, the {\it retweet}, the {\it quote} and the {\it reply}. We will focus on the {\it mentions} and the {\it  retweets}, as they are the most widely used. 

The {\it mention} is a public direct communication mechanism. It consists in including the name of a user in a {\it tweet} ([..]@username[...]). This way, the mentioned user will receive the message even if she is not following the user that posted it. There is a convention in Twitter according to which, whenever a given user is mentioned inside a tweet, the {\it mention} mechanism is used, independently of whether the intention is to establish a communication with the target user or not. 

The {\it retweet} is a broadcast mechanism. Every retweeted message by a user is forwarded to her followers the same way as an original tweet. In a political context retweeting usually, but not always, implies the endorsement of the ideas contained in the message.

\subsection{Building the networks of mentions and retweets}
\label{sec_build_net}

Here we will explain the methodology that we have adopted to build the networks of interactions from Twitter data. The interactions that we have analyzed are the {\it mention} and the {\it retweet}.

We have built mention networks by considering each user participating in the conversation as a node. Two nodes {\it j} and {\it i} are joined with a directed link from $j$ to $i$ when user {\it j} mentions user {\it i}; that is, {\it j} posts a {\it tweet} including {\it @name\_of\_i}. The result is a directed and weighted network where the weight of each link corresponds to the total number of tweets where {\it j} mentions {\it i} gathered during an given interval of time. The in-degree, or $k_{in}^i$, of a user $i$ in this network corresponds to the total number of users that have mentioned $i$. The $k_{out}^i$ corresponds to the total number of users that she has mentioned.

\begin{figure}[htbp]
  \begin{center}
  \includegraphics[width=0.45\textwidth]{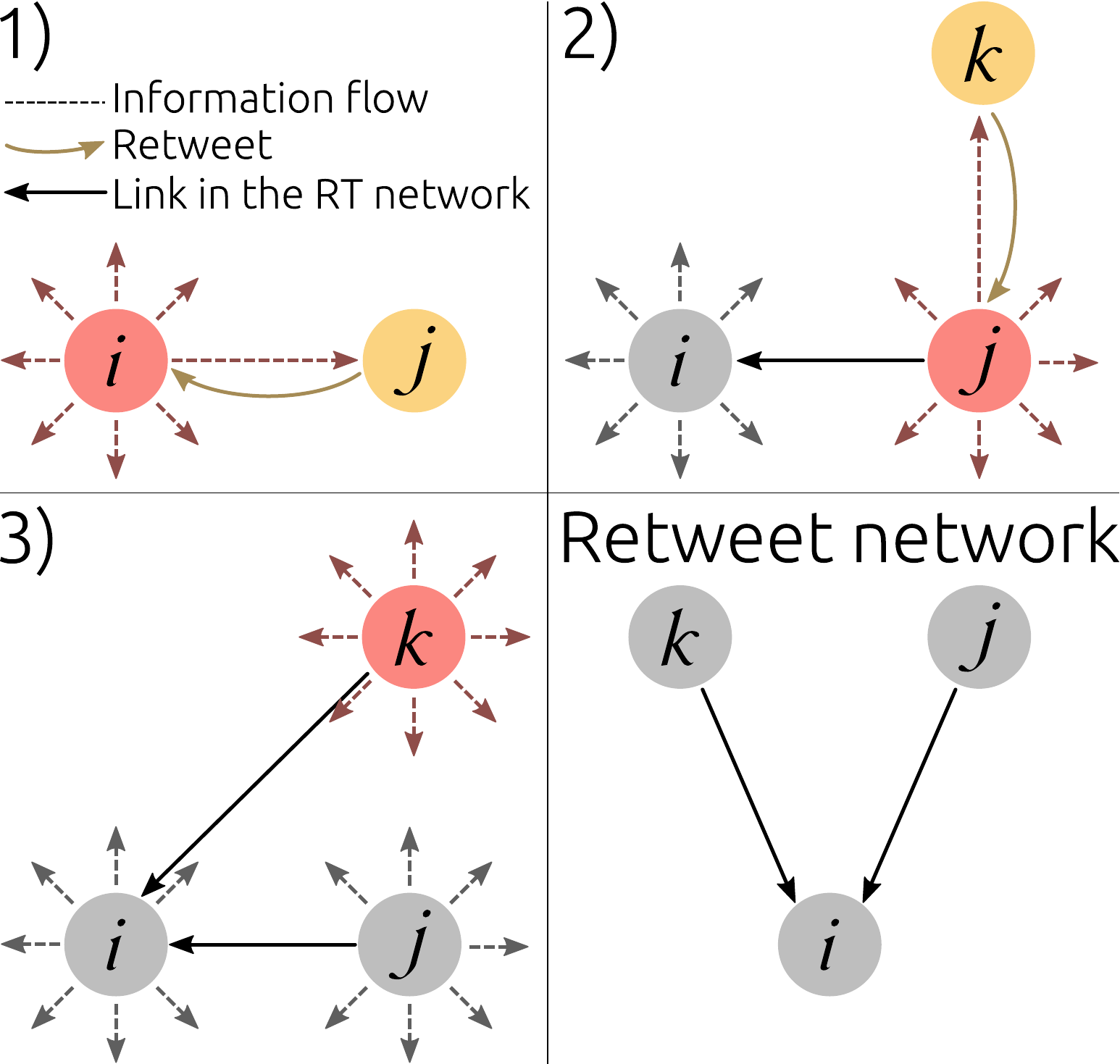}
  \end{center}	
  \caption{Scheme that shows the method used to build the networks of retweets. In panel 1), user $i$ posts a message starting an information flow that reaches user $j$, who decides to retweet it. In panel 2), the retweeted message posted by $j$ is read by $k$ who in turn retweets it again. In panel 3) the three users are information transmitters and in the last panel the resulting retweet network is shown.}
  \label{fig_rt_scheme} 
  \end{figure}

The retweet network is built in a similar way as the mention network. The nodes are the users and user {\it j} is joined with a directed link to user {\it i} if {\it j} retweets a message {\it originally} posted by {\it i}. If a user $j$ retweets an original tweet by $i$ and $k$ retweets the message broadcast by $j$, there are links from $k$ to $i$ and from $j$ to $i$, but not from $k$ to $j$. Figure \ref{fig_rt_scheme} illustrates this mechanism. 

Let us remark this idea: links in the retweet network join retweeters with original posters; there are no links between the middlemen that broadcast the original message. The reason for this choice of methodology is that the Twitter API used to download the data at that moment only provided information about the original poster of a tweet in the retweet metadata. The retweet network is then a directed and weighted network where the weights of the links from $j$ to $i$ are the number of messages of {\it i} retweeted by {\it j} during a given period of time. The $k_{in}^i$ of a user $i$ in this network corresponds to the total number of users that have retweeted messages by $i$. The $k_{out}^i$ corresponds to the total number of users whose messages she has retweeted. The retweet mechanism usually results in networks that display large star subgraphs.

We have considered two different temporal scales. On one hand, we have built daily networks, counting the 24 hours of a day from 4AM (UTC) \footnote{In Spain, the local time corresponds to UTC+1 for winter time and UTC+2 for summer time.} of that day until 4AM of the following day. This way we capture full {\it human} activity cycles. This approach is based on a work by Morales et al. \cite{morales20161048}, where human synchronization is studied at several scales, finding strong periodicities linked to day-night and social cycles as well as a developing global synchronization that adds a new perspective to the phenomenon of globalization. On the other hand, we have built aggregated networks for the 15 days of electoral campaign plus the next three days (pre-election, election and post-election).

\section{Results}

\subsection{Temporal evolution of user activity}
\label{sec_user_act}

In order to characterize the global activity of the users (number of tweets posted during a given period of time), we have computed time series of total daily activity for the whole period considered. Additionally, we have analyzed the temporal evolution of the distribution of daily activity per user.

\begin{figure*}[htbp]
  \begin{center}
  \includegraphics[width=0.95\textwidth]{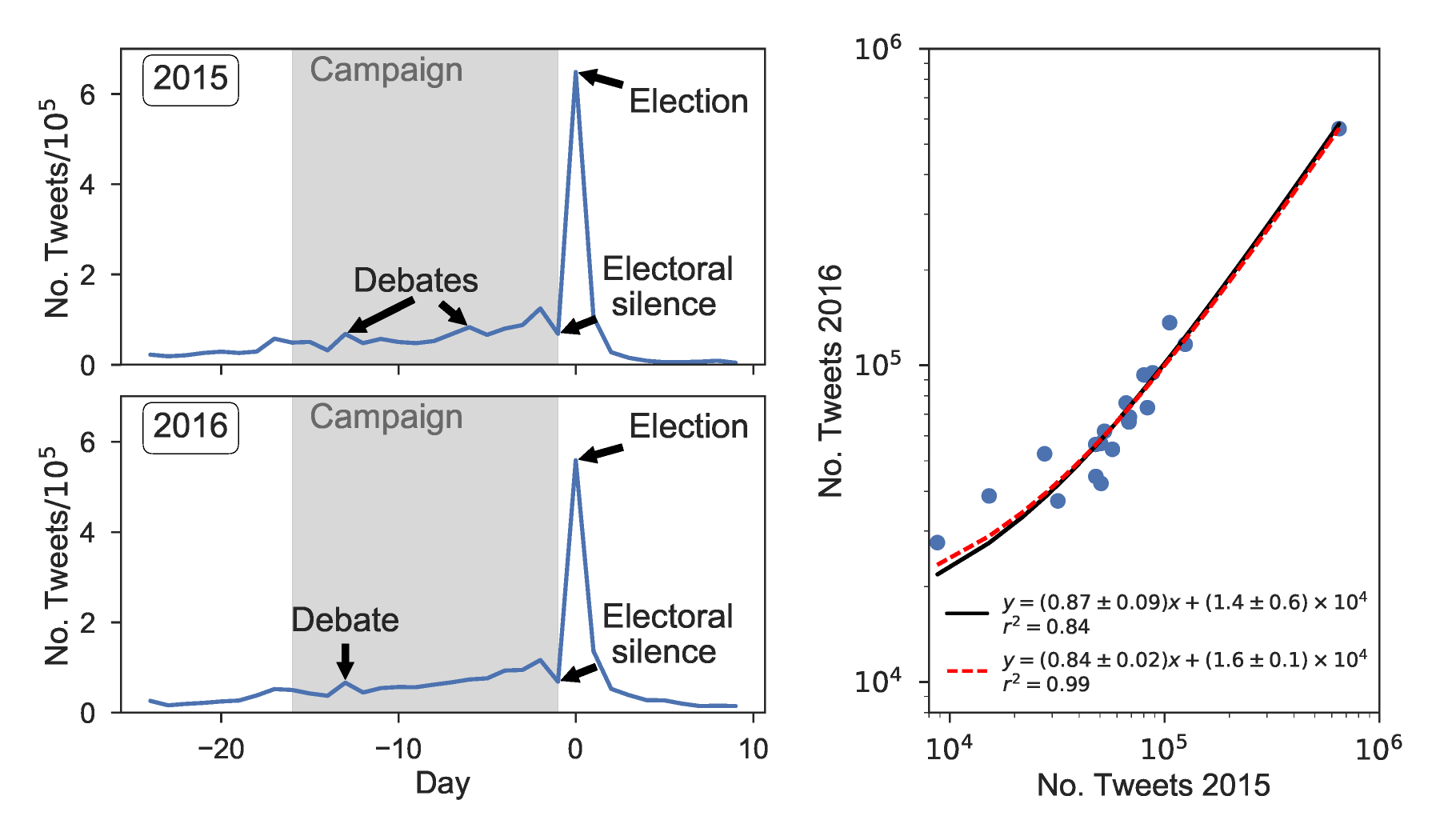}
  \end{center}	
  \caption{Left panels: time series of aggregated user activity per day for the 2015 (top) and 2016 (bottom) elections. The shadowed region corresponds to the days of electoral campaign. Right panel: Linear regression between the activity time series of both elections. The lines correspond to the linear fits of the data including the day of the elections (red dashed line) and excluding it (black continuous line).}
  \label{fig_ts_act}
  \end{figure*}

In the left panels of figure \ref{fig_ts_act} we have plotted the time series of daily activity for both electoral campaigns. We have aligned them such that the voting day coincides with the day 0. As we can see in that figure, before the electoral campaign the time series of daily activity present low values, but they start to rise around the onset of both campaigns.

In line with the existing literature \cite{doi:10.1063/1.4729139,doi:10.1177/1461444811422894,Jungherr:2013:TVS:2508436.2508437}, the features of the time series of activity can be explained to a high extent by offline events. There is a huge peak of messages on the election day, which accounts for more than the 30\% of the total of messages during the period of study (36\% in 2015 election and 32\% in 2016). There are also smaller peaks during the period of the electoral campaign that can be explained by the main electoral debates. 
 
 In Spain, pre-electoral silence is mandatory during the period that spans from 12AM of the day before the election to the closing of voting polls. This period is called {\it reflection day} and a number of restrictions apply with respect to the behavior of political parties and media. Among others, electoral events can not be held and the spread of electoral propaganda is forbidden \cite{leyelectoral}. The time series experiences a decrement that day, which may be triggered by the fact that politicians cease to campaign in both online and offline media. Because of this lack of stimulous, the conversation among common users decay.

It is worth pointing out the similarity of the temporal evolution of the user activity in both elections. In figure \ref{fig_ts_act}, we have plotted (in logarithmic scale) the time series of one period against the other such that the election days coincide. By performing a linear regression of these data, a significant linear correlation has been found. We obtain an $r^2=0.99$ when we perform the linear regression including the day of the elections and an $r^2=0.84$ if we exclude it. However, notice that the slopes present only a difference of $3.5\%$. This correlation can be partially explained by the similarity of the contexts and by the recurrent structure of the electoral campaign: It always starts 16 days before the election and lasts 15 days. Moreover, the main debates were both held 13 days before the election. Due to the fact that the number of tweets in the 2016 election was slightly lower than in 2015, the slope of the linear regression is lower than 1.

\begin{figure}[htbp]
  \begin{center}
  \includegraphics[width=0.45\textwidth]{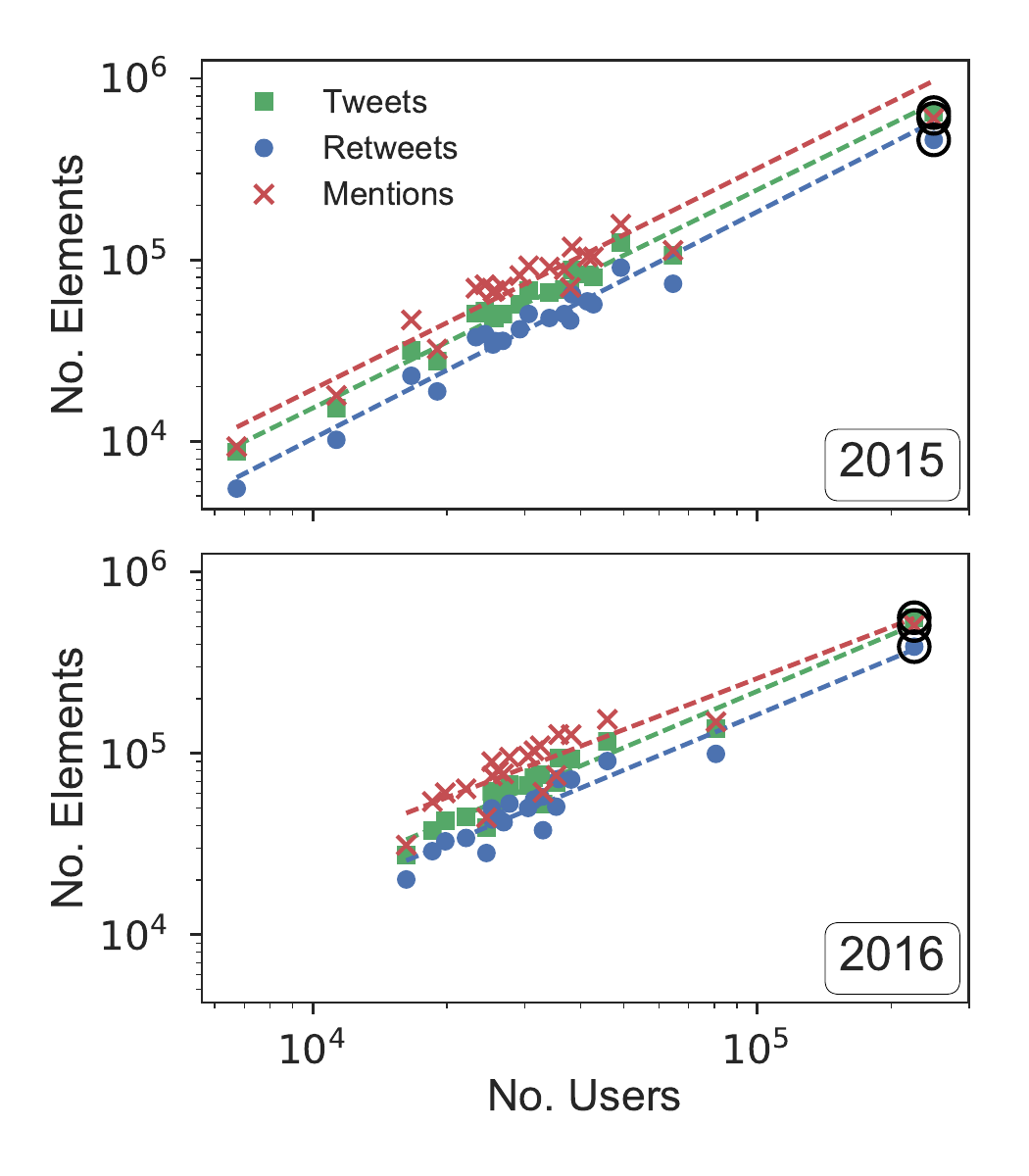}
  \end{center}
  \caption{Power law relationships of the total number of tweets, retweets and mentions per day as a function of the number of unique users that participated in the conversation each day for the 2015 campaign (top) and the 2016 campaign (bottom). Note that the data corresponding to the day of the elections (marked with a circle) were not included in the fit.}
  \label{fig_lin_rels}
  \end{figure}

Another relevant property of the user activity is displayed in figure \ref{fig_lin_rels}, where we show that the total number of tweets, retweets and mentions per day follow a power law with respect to the number of unique users each day. Accordingly, we have fitted our data to a power law with exponent $\alpha$. The values of the exponent $\alpha$ and correlation coefficients of the fits are displayed in table \ref{tab_lin_rels}. To understand this behavior we have followed the work by Leskovec et al. \cite{leskovec2005graphs}, where it is shown that when real-world networks evolve through time, the number of links $E(t)$ scale with the number of nodes $N(t)$ as:

\begin{equation}
E(t) \propto N(t)^a
\end{equation}

Notice that, whereas in 2015 the growth for the three quantities was slightly super-linear with respect to the number of users, in 2016 we observe an approximately linear behavior. Hence, in 2015 when more users join the conversation, the activity experiences a proportionally higher increment than in 2016.

\begin{table}[htbp]
\caption{Values of the exponents $(\alpha)$ and correlation coefficients ($r^2$) of the linear regressions of the log-log plots of tweets, retweets and mentions with respect to the number of unique users per day shown in figure \ref{fig_lin_rels} for both electoral campaigns.}

\begin{tabular}{|l|c|c|c|c|}
\hline
\multicolumn{ 1}{|c|}{\textbf{}} & \multicolumn{ 2}{c|}{\textbf{$\alpha$}} & \multicolumn{ 2}{c|}{\textbf{$r^2$}} \\ \cline{ 2- 5}
\multicolumn{ 1}{|l|}{} & \textbf{2015} & \textbf{2016} & \textbf{2015} & \textbf{2016} \\ \hline
Tweets & $1,20 \pm 0,06$ & $1,04 \pm 0,11$ & 0,96 & 0,85 \\ \hline
Retweets & $1,25 \pm 0,08$ & $1,02 \pm 0,12$ & 0,94 & 0,80 \\ \hline
Mentions & $1,22 \pm 0,10$ & $0,9 \pm 0,2$ & 0,89 & 0,66 \\ \hline
\end{tabular}

\label{tab_lin_rels}
\end{table}

In order to further explore the characteristics of the user behavior, we have also analyzed the temporal evolution of the distribution of the daily user activity shown in the left panels of figure \ref{fig_distrib_act}. The activity distributions show a heterogeneous character, result that is in concordance with the literature \cite{Morales20141}. We have fitted them to the following discrete power law:

\begin{figure*}[htbp]
  \begin{center}
  \includegraphics[width=0.95\textwidth]{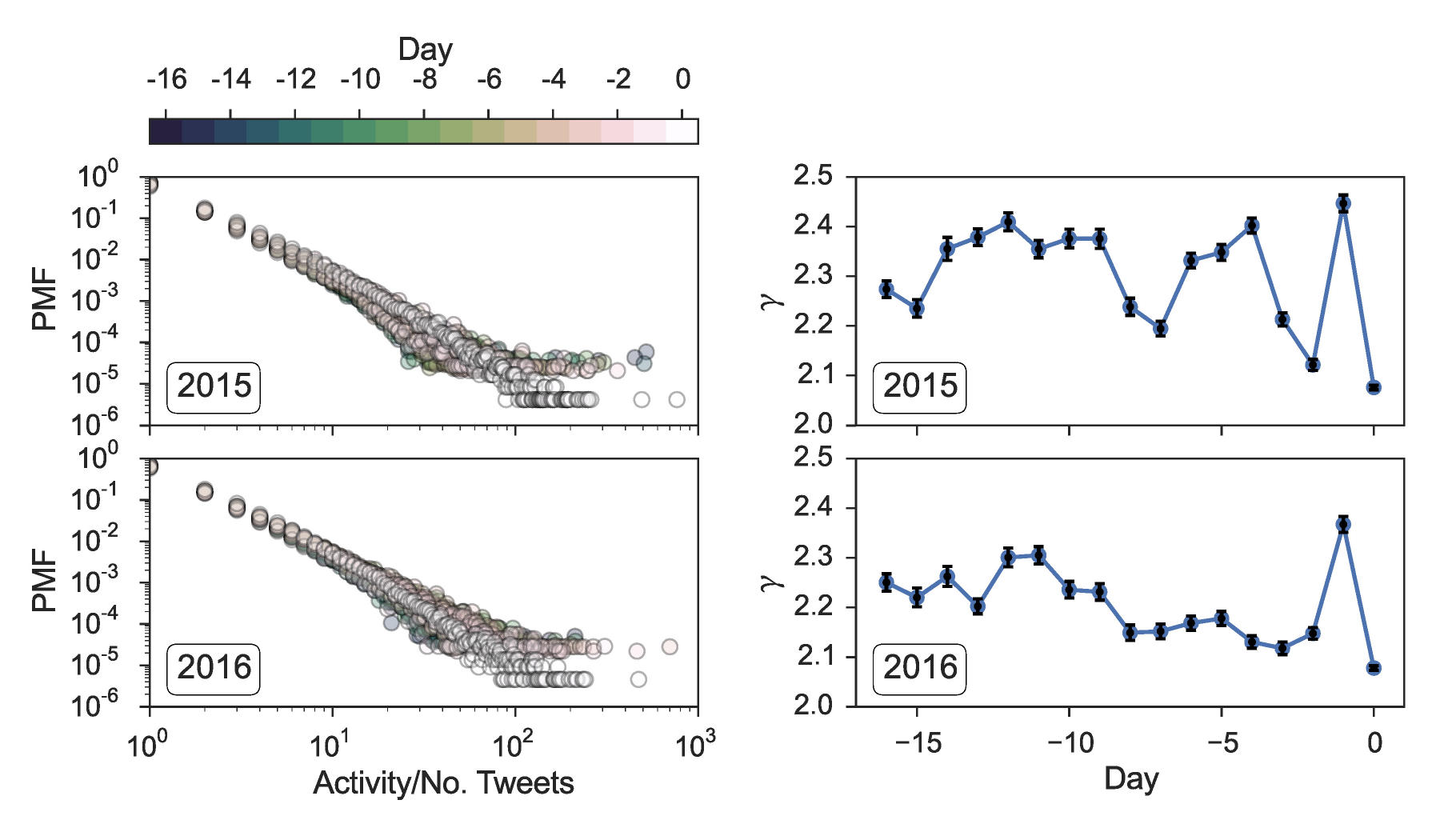}
  \end{center}	
  \caption{Temporal evolution of the distribution of activity per day in both electoral campaigns (2015 in the top panels and 2016 in the bottom panels). Left panels: probability mass functions (PMF) of the distribution of activity for each day in color code. Right panels: daily evolution of the $\gamma$ exponent of the power law fit of the activity distributions. The error bars correspond to $2 \sigma$.}
  \label{fig_distrib_act}
  \end{figure*} 

\begin{equation}
P(x) = \frac{x^{-\gamma}}{\zeta(\gamma,x_{min})} \quad ; \quad \zeta(\gamma,x_{min}) = \sum_{x=0}^{\infty} \frac{1}{(x+x_{min})^{\gamma}}
\end{equation}

Where in this case $x$ is the daily activity and $\zeta$, the Riemann zeta function.

 In the right panels of figure \ref{fig_distrib_act} we have plotted the temporal evolution of the exponent $\gamma$ of the power law. It has been computed by numerically solving the following equation  to obtain the maximum likelihood estimator \cite{doi:10.1137/070710111}  (MLE) of $\gamma$:
 
\begin{equation}
 \frac{\zeta'(\gamma,x_{min})}{\zeta(\gamma,x_{min})} = - \frac{1}{n}\sum_{i=1}^{n} \ln (x_i)
 \end{equation}
 
 Where the prime denotes differentiation with respect to the first argument and $n$ corresponds to the number of users. The uncertainty of $\gamma$ has been computed as:
 
 \begin{equation}
 \sigma = \frac{1}{\sqrt{n \left[  \frac{\zeta''(\gamma,x_{min})}{\zeta(\gamma,x_{min})} - \left( \frac{\zeta'(\gamma,x_{min})}{\zeta(\gamma,x_{min})} \right)^2 \right]}}
 \end{equation}
 
 This technique is proven to be more precise than a minimum squares fit to the log-log plot, which usually yields incorrect results when computing the parameters of power law distributions\cite{doi:10.1137/070710111}. Every other exponent of power laws of discrete data has been computed in the same way. Since the minimum value of activity is 1 (and is also the most abundant in the data), we have fixed $x_{min}=1$ when computing the MLE of $\gamma$.
 
We can see that the values of the $\gamma$ exponent fluctuate within a small interval ($2.0 < \gamma < 2.5$) during the considered period for both elections. These values are perfectly compatible with those presented in \cite{doi:10.1063/1.4729139}, where the authors obtain a value of $\gamma = 2.275 \pm 0.002$.

The small fluctuations of the $\gamma$ exponent tell us that the collective behavior does not change much from one day to the other. If we take into account that the value of the exponent of the distribution controls how fast it decays, we see that, when the exponent is smaller, the activity reaches higher values (and vice versa). Consequently, by looking at figure \ref{fig_distrib_act}, we can see that the day of the elections we obtain a low exponent due to the increase in activity. Additionally, during the electoral silence, consistently with the results shown in figure \ref{fig_ts_act}, we appreciate a decrease in activity. Notice however that, while in the case of figure \ref{fig_distrib_act} the points corresponding to the electoral silence are the highest, they are not the lowest in the activity time series of figure \ref{fig_ts_act}. This implies that, although individually each user tends to post fewer messages that day, there are still a lot of users taking part in the conversation.

\subsection{Temporal evolution of mention and retweet networks}

We have analyzed the temporal evolution of the aggregated mention and retweet networks at two different temporal scales. On one hand, we have aggregated the networks for the whole campaign period (plus the next three days); on the other hand, we have performed an analysis of the temporal evolution of the networks by aggregating the data for each day separately and computing time series for different metrics. 

\begin{figure*}[htbp]
  \begin{center}
  \includegraphics[height=0.3\textheight]{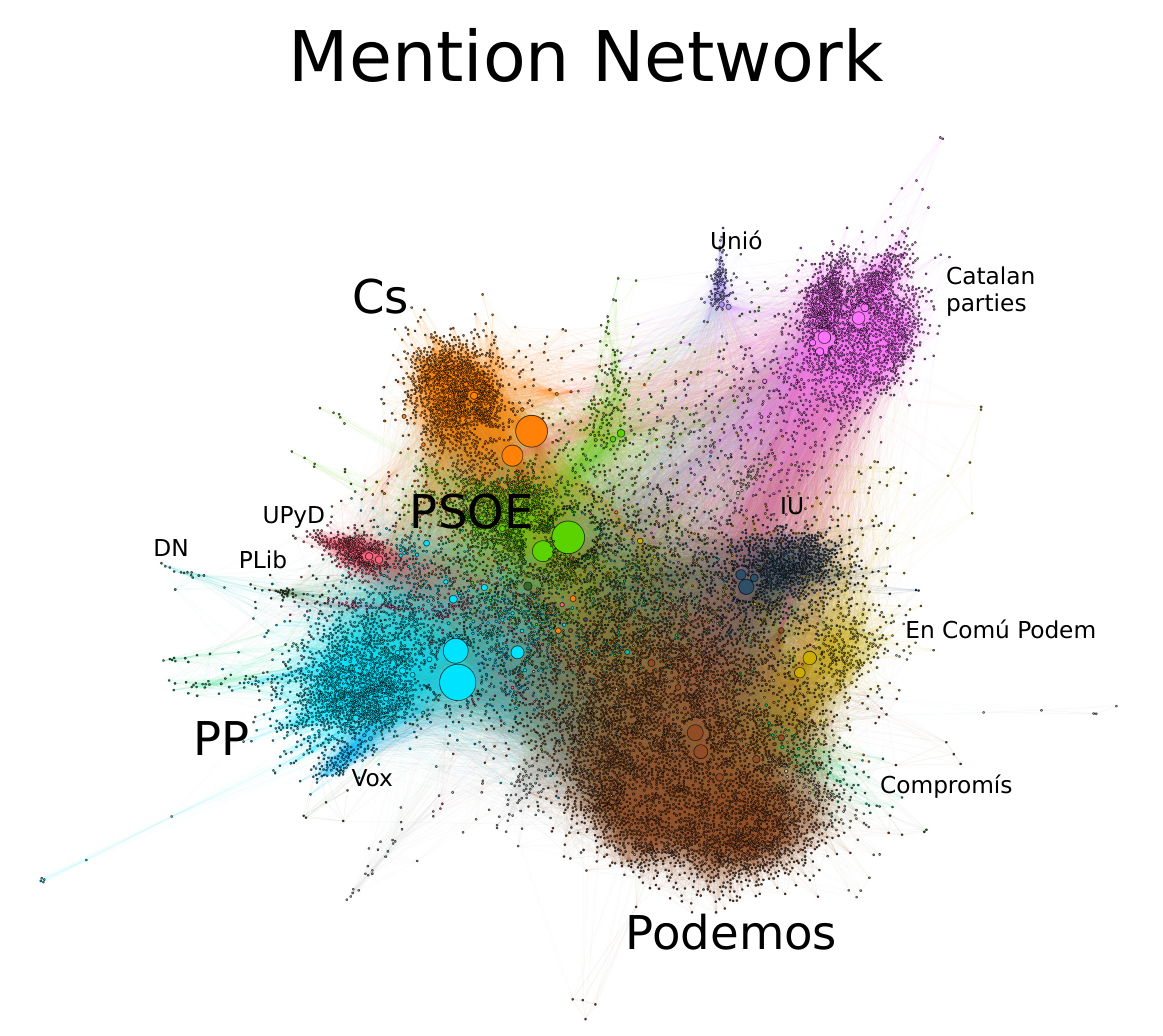}
    \includegraphics[height=0.295\textheight]{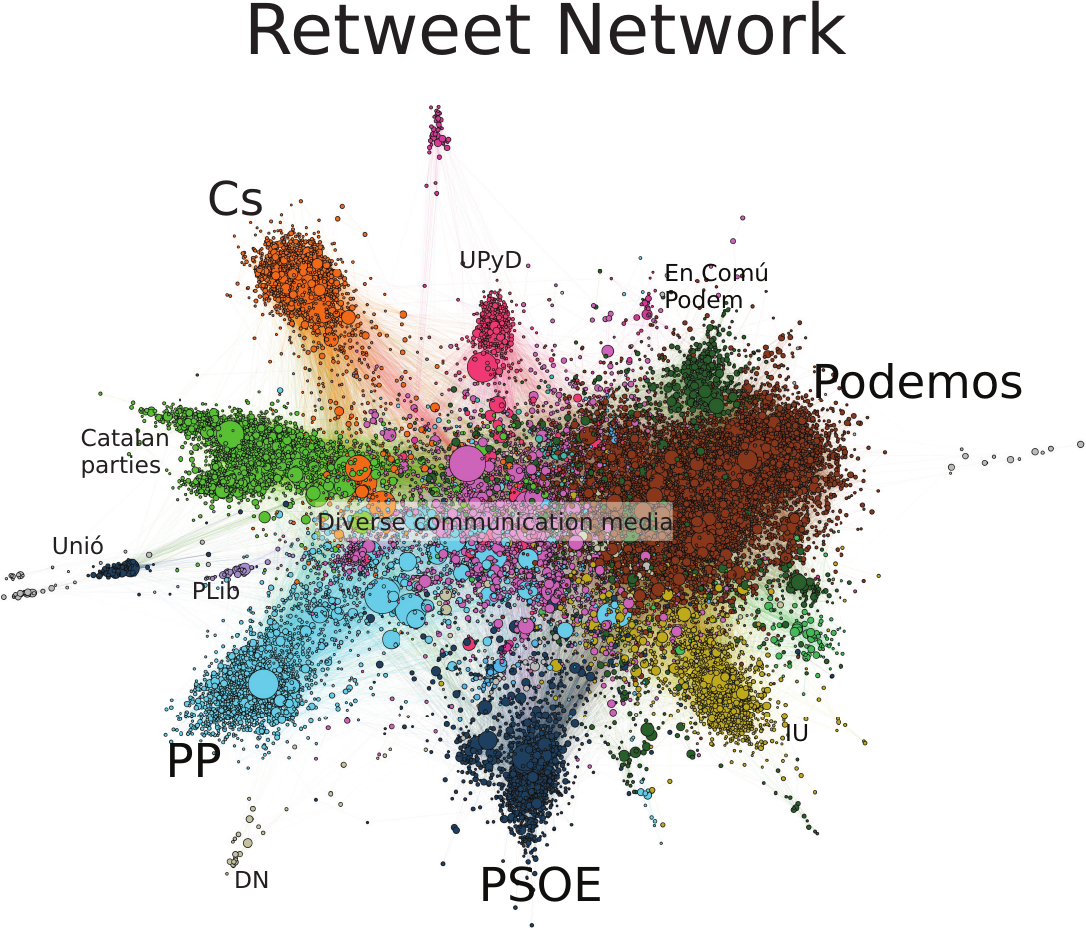}
  \end{center}
  \caption{Left panel: strongly connected component of the aggregated mention network for the 2015 electoral campaign. Right panel: strongly connected component of the aggregated retweet network for the 2015 electoral campaign. Colors correspond to the communities computed with the Louvain algorithm \cite{1742-5468-2008-10-P10008}. We have indicated the most probable affiliations of the nodes of each community by visually inspecting which users correspond to the most central nodes. Every well defined group seems to correspond to a political party. The size of the nodes is proportional to $\log(PageRank)$.}
  \label{fig_nws}
  \end{figure*}

In figure \ref{fig_nws} we show the strongly connected component of the aggregated networks of 2015. Colors correspond to the communities computed with the Louvain algorithm \cite{1742-5468-2008-10-P10008}.  We have obtained a modularity of $Q=0.72$ for the retweet network and $Q=0.67$ for the mention network. Other community detection algorithms were applied obtaining analogous results \cite{PhysRevX.4.011047}. We have indicated the most probable affiliations of the nodes of each community. In order to do that we have visually inspected which nodes (or users) are the most central in each community. The centralities of the nodes have been computed with the $PageRank$ metric 	\cite{brin1998anatomy}, which measures the influence of a node based on its neighborhood. The sizes of the nodes in figure \ref{fig_nws} have been represented proportional to $\log(PageRank)$.

Every well defined group seems to correspond to a political party. Whereas in the representation of the mention network the different groups are well defined and the most central nodes correspond to the leaders of the communities and are politicians or political parties, in the retweet network a mixing of nodes of different communities placed in the center of the representation can be appreciated. Most of these nodes present high centralities and belong to different communication media. This is in good agreement with the literature \cite{borondo201590}.

We have computed several statistical properties of these networks: the number of nodes ($N$) and links ($E$), the density ($\rho=\frac{E}{N(N-1)}$) of the network and the in and out degree distributions. We have found that they are very heterogeneous and fitted them to a power law following the methodology described in section \ref{sec_user_act}. The in-degree ($k_{in}$) distribution is more heterogeneous than the out-degree ($k_{out}$) distribution; that is, the $\gamma$ exponents follow the relation $\gamma_{out} > \gamma_{in}$. This is due to the fact that the $k_{out}$ is associated to individual efforts whereas the $k_{in}$ corresponds to collective actions. An individual is physically limited to posting a given number of tweets during a given period of time, whereas a large group of users may post many more messages in the same period of time. Because of this, high $k_{in}$ values are more probable than high $k_{out}$ values, leading to the aforementioned relationship between the $\gamma$ exponents

The properties of the aggregated networks are displayed in table \ref{tab_gen_stat_nw}, where we can appreciate clear differences between the mention and the retweet networks. First of all, the former presents higher average and maximum degree. Whereas the higher $\gamma_{in}$ of the mention network shows that its $k_{in}$ distribution is slightly less heterogeneous than the one for the retweets, the lower value of $\gamma_{out}$ implies that the $k_{out}$ distribution is more heterogeneous for mentions than for retweets. The first phenomenon can be attributed to the construction methodology of the networks. Since all retweeters are linked to the original poster, high values of $k_{in}$ are slightly more common in the retweet network. The second phenomenon arises from the fact that there can be more than one mention per tweet and in every retweet there is, at least, one mention to the original poster.

The higher average clustering $(\bar{C})$ for mention networks can be explained by the different uses of mentions and retweets. Mentions are a communication and allussion mechanism whereas the retweets are used mainly to broadcast messages, leading to more clustered networks on the first case and to more star-like subgraphs on the second one. Nevertheless, the construction mechanism of the networks also plays a relevant role, since the relationships between middlemen are not present in the retweet network.

\begin{table}[htbp]

\caption{General statistical properties of the retweet and mention networks aggregated for the whole period of study for both electoral campaigns. $k_{in(out)}$ corresponds to in (out) degree and $\gamma_{in(out)}$ to the exponent of the degree distribution. The confidence intervals in the power law exponents calculations correspond to $2\sigma$. }
\resizebox{0.49\textwidth}{!}{
\begin{tabular}{|l|c|c|c|c|}
\hline
\multirow{2}{*}{\textbf{Statistics}} & \multicolumn{ 2}{|c|}{\textbf{Mention}} & \multicolumn{ 2}{c|}{\textbf{Retweet}} \\ \cline{2-5}
 & \textbf{2015} & \textbf{2016} & \textbf{2015} & \textbf{2016} \\  \hline
Nodes & 354079 & 319961 & 330071 & 296645 \\  \hline
Links & 1361495 & 1251113 & 928492 & 852216 \\  \hline
Density & 1,09E-05 & 1,22E-05 & 8,52E-06 & 9,68E-06 \\  \hline
$\bar{k_{in}} (=\bar{k_{out}})$ & 3,85 & 3,91 & 2,81 & 2,87 \\  \hline
$(k_{in}^{min},k_{out}^{min})$ & (0,0) & (0,0) & (0,0) & (0,0) \\  \hline
$(k_{in}^{max},k_{out}^{max})$ & (17747, 824) & (16829, 645) & (9054, 459) & (8985, 534) \\  \hline
$\gamma_{in}$ & $1,606 \pm 0,005$ & $1,623 \pm 0,005$ & $1,587 \pm 0,005$ & $1,577 \pm 0,006$ \\  \hline
$\gamma_{out}$ & $1,827 \pm 0,003$ & $1,825 \pm 0,003$ & $2,034 \pm 0,004$ & $2,022 \pm 0,004$ \\ \hline
$\bar{C}$ & 0.19 &0.19 & 0.06& 0.06\\  \hline
\end{tabular}
}
\label{tab_gen_stat_nw}
\end{table}

Nevertheless, the results show very little change from one election to the other, suggesting that the underlying dynamics of the networks of interactions are consistent and, to some extent, independent of the context.

\begin{figure}[htbp]
  \begin{center}
  \includegraphics[width=0.47\textwidth]{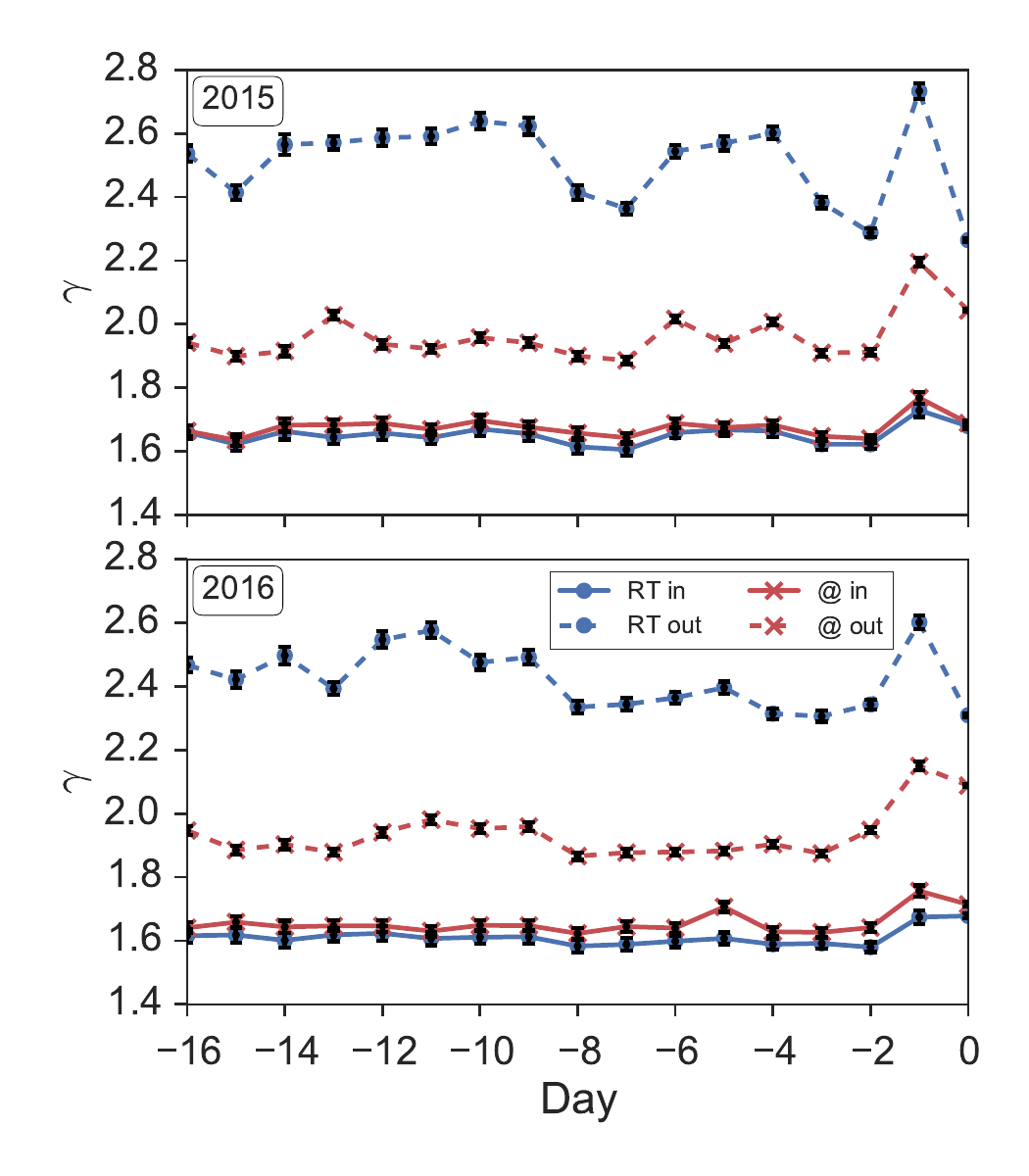}
  \end{center}	
  \caption{Daily evolution of the $\gamma$ exponent of the in and out degree distributions for mentions (@) and retweets (RT) networks for the 2015 (top) and 2016 (bottom) electoral campaigns.}
  \label{fig_inout_degexp}
  \end{figure}

With respect to the temporal evolution of the exponent of the in-degree distribution for mention and retweet networks, which is shown in figure \ref{fig_inout_degexp}, it remains approximately constant for 47 days before the election (for simplicity, only 16 days before the election are shown in the figure). The average values during that period are $\gamma_{in}^{2015}=1.68\pm0.08$ and $\gamma_{in}^{2016}=1.67\pm0.07$ for the mention networks and $\gamma_{in}^{2015}=1.65\pm0.08$ and $\gamma_{in}^{2016}=1.63\pm0.07$ for retweet networks, with a confidence interval of $2\sigma$. These results indicate a high consistency of the user behavior for both elections. The exponents are slightly higher to the one of the aggregated networks because in those networks almost everyone has more mentions, making the tail of the distribution longer and heavier.

In the case of the temporal evolution of the out-degree distribution exponent, displayed in figure  \ref{fig_inout_degexp}, it also  remains approximately constant for 47 days before the election. The average values during that period are $\gamma_{out}^{2015}=1.99\pm0.09$ and $\gamma_{out}^{2016}=2.00\pm0.08$ for the mention networks and $\gamma_{out}^{2015}=2.7\pm0.3$ and $\gamma_{out}^{2016}=2.6\pm0.2$ for the retweet networks, with a confidence interval of $2\sigma$. Analogously to the in-degrees, the similarity of the values of the exponents from one year to the other suggests a recurrent behavioral pattern. They are also similar to the values of the aggregated network, but higher, meaning that the heterogeneity is lower. The reason is again the construction mechanism of these networks. People are active during all the period under study, producing more links each day, thus slightly changing the distribution to a more heterogeneous one as more days are aggregated.

The fact that the exponents of the out-degree distributions remain approximately constant (see figure \ref{fig_inout_degexp}) is deeply related with the linear relationship between retweets and mentions with respect to the number of unique users discussed in section \ref{sec_user_act}.

We have explored the degree correlations of the networks and their evolution in several ways. First, we have computed the degree assortativities \cite{Foster15062010}, presented in table \ref{tab_assort_nw}, and compared them with assortativities of randomized networks via a Z-Score in order to determine if the assortativities are a result of the degree distributions alone or of more complex correlations. The Z-Scores can be consulted in table \ref{tab_ZSassort_nw}. The randomization have been carried out using 500 realizations of the directed configuration model \cite{PhysRevE.64.026118} implemented in the NetworkX Python module \cite{hagberg-2008-exploring}.

\begin{table}[htbp]
\caption{Assortativities of the different aggregated networks.}

\begin{tabular}{|l|c|c|c|c|}
\hline
\multirow{2}{*}{\textbf{Link direction}} & \multicolumn{ 2}{c|}{\textbf{Mention}} & \multicolumn{ 2}{c|}{\textbf{Retweet}} \\ \cline{ 2- 5}
 & \textbf{2015} & \textbf{2016} & \textbf{2015} & \textbf{2016} \\ \hline
out-in & -0,1096 & -0,1095 & -0,1234 & -0,1169 \\ \hline
in-in & -0,0174 & -0,0162 & -0,0313 & -0,0287 \\ \hline
in-out & -0,0006 & -0,0014 & 0,0109 & 0,0074 \\ \hline
out-out & 0,0521 & 0,0488 & 0,0988 & 0,0892 \\ \hline
\end{tabular}

\label{tab_assort_nw}
\end{table}

\begin{table}[htbp]
\caption{Z-Scores of the assortativities of the different aggregated networks with respect to 500 realizations of the directed configuration model.}

\begin{tabular}{|l|c|c|c|c|}
\hline
\multirow{2}{*}{\textbf{Link direction}} & \multicolumn{2}{c|}{\textbf{Mention}} & \multicolumn{2}{c|}{\textbf{Retweet}}  \\ \cline{2-5}
 & \textbf{2015} & \textbf{2016} &\textbf{2015}&\textbf{2016} \\ \hline
out-in& -152 & -137 & -141 & -133 \\ \hline
in-in & -22 & -19 & -31 & -31 \\ \hline
in-out & 2 & 2 & 12 & 9 \\ \hline
out-out & 100 & 95 & 107 & 113 \\ \hline
\end{tabular}

\label{tab_ZSassort_nw}
\end{table}

As we can see in table \ref{tab_assort_nw}, the assortativities display a recurrent order among networks and years: $r_{out-in} < r_{in-in} < r_{in-out} (\approx 0) < r_{out-out}$. Consistently with the rest of the measures presented so far, they exhibit a high similarity among years. Moreover, when we compare the results of the degree correlations of the retweet network to those of the 2011 election \cite{doi:10.1063/1.4729139}, we see that they are very similar. This result suggest a recurrent behavior in the communication patterns of the users.

This order is maintained for the Z-Scores (see table \ref{tab_ZSassort_nw}), meaning that the assortative relations are still more assortative than what we should expect for a random network and that the disassortative ones are more disassortative.

\begin{figure}[htbp]
  \begin{center}
  \includegraphics[width=0.47\textwidth]{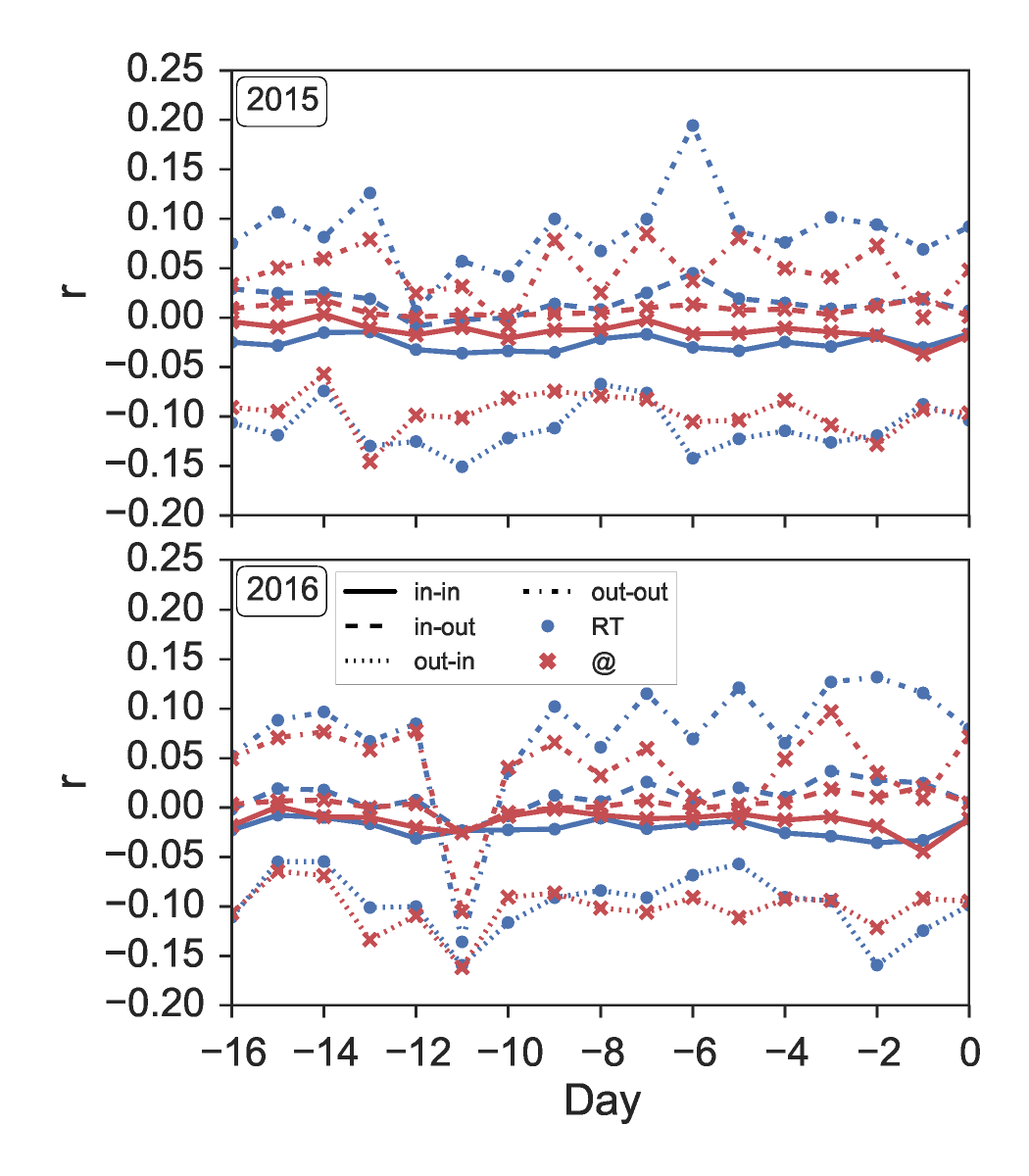}
  \end{center}	
  \caption{Daily evolution of the directed degree assortativities  for mentions (@) and retweets (RT) networks for the 2015 (top) and 2016 (bottom) electoral campaigns.}
  \label{fig_deg_assort}
  \end{figure}

With respect to the temporal evolution of the daily degree assortativities that is displayed in figure \ref{fig_deg_assort}, we can see that they fluctuate around a more or less fixed value. However, these fluctuations are sometimes strong enough to alter the previously described order. 

These fluctuations seem to be caused by disruptive events, both exogenous and endogenous. On day -6 of 2015 a debate was celebrated between the leader of the PP (then, the president) and the leader of the PSOE. A similar pattern, but weaker, can be observed in the time series of assortativity on day -13, which coincides with the other important debate of the campaign. On day -11 of 2016 we have found a viral tweet that contained a comical video about Spanish politics. The retweets received by that particular tweet amounted to $4,7\%$ of all the tweets published that day in the analyzed conversation, when the average proportion of retweets received by the most popular tweets each day during the 2016 campaign was $(1,7 \pm 0,3)\%$.

\subsection{Communication efficiency}

In order to measure the global influence of a user on the network, we have used the user efficiency metric \cite{Morales20141}. In Twitter, it can be considered that a message has reached a higher impact if it gets a high number of retweets (RTs). A user is more efficient the higher her average number of RTs per original tweet is. Hence, the efficiency of a user is defined as follows:

\begin{equation}
\eta_i = \frac{RTs}{Posts} = \frac{(s_{in}^{RT})_i}{A_i}
 \end{equation}
 
 Where $(s_{in}^{RT})_i$ is the in-strength of the user $i$ in the retweet network and $A_i$ is her activity, that is, the total number of posted messages. It should be noted that $\eta \in [0, \infty)$. Therefore, this measure represents the collective response to the individual action. 
 
 We have computed the distribution of efficiency for the global conversation including regular users and the politicians that participated in it. In order to compare the behavior of regular users with politicians, we have also considered the distribution of efficiency corresponding to the groups of accounts of the four main political parties described in section \ref{sec_data}.

\begin{figure}[htbp]
  \begin{center}
  \includegraphics[width=0.47\textwidth]{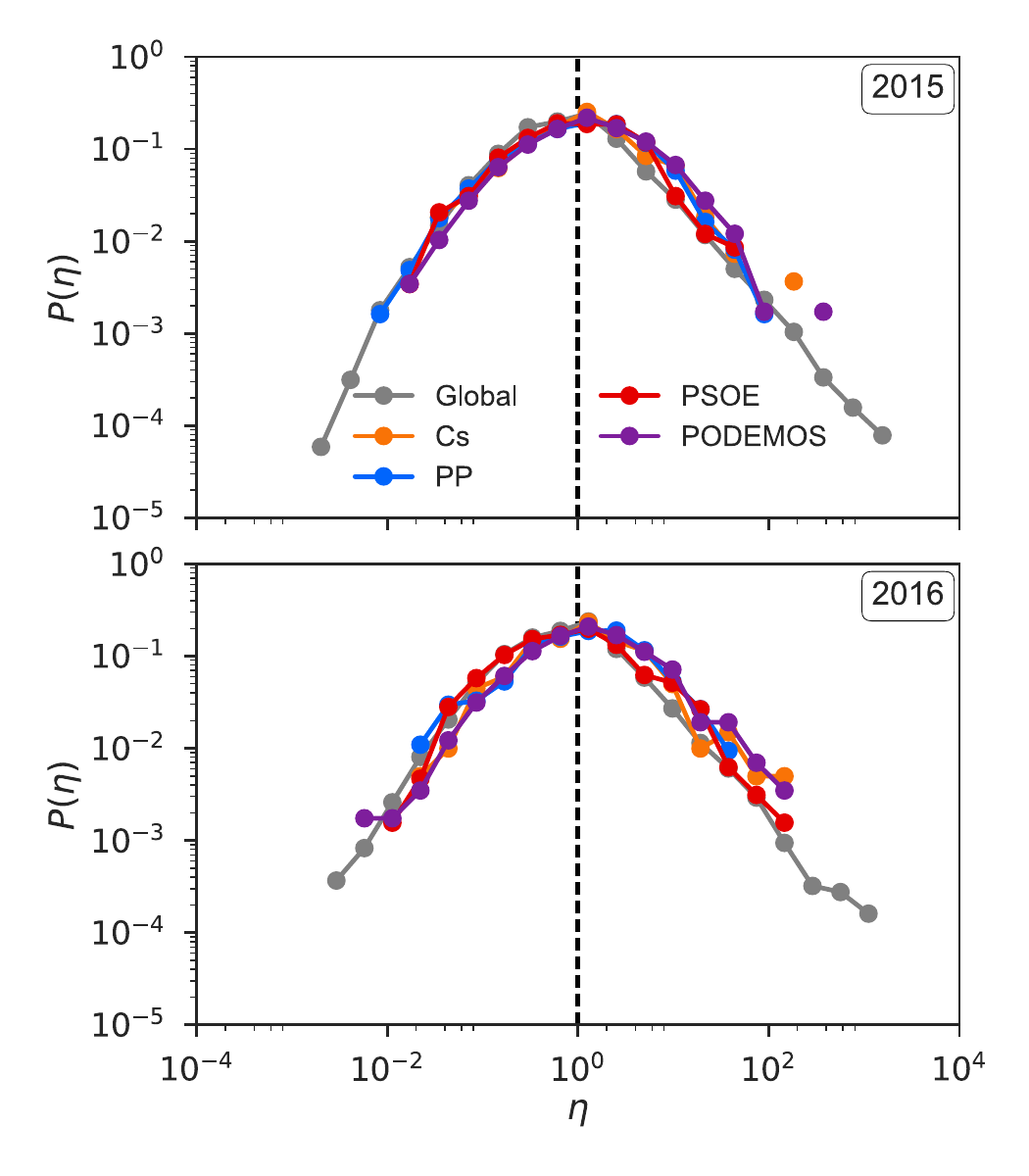}
  \end{center}
  \caption{Probability distributions of efficiency for Twitter accounts associated to each party and for the whole set of users. Top: 2015 electoral campaign. Bottom: 2016 electoral campaign. The value $\eta=1$ is marked with a dashed line.}
  \label{fig_distr_eff}
  \end{figure}

In figure \ref{fig_distr_eff} we represent, for both electoral campaigns, the probability distributions of efficiency for Twitter accounts associated to each party and for the whole set of users. This distribution corresponds to the probability mass function.
 There, we can appreciate that the distributions follow the same functional form for common users and users associated to political parties.  In particular, they show a heterogeneous behavior characterized by a heavy tail with power law decay. This has been tested by fitting the tail of the distribution to different functions (exponential, gamma, lognormal and power law) and choosing the best one according to the Akaike Information Criterion (AIC)\cite{1100705}. The transition to the power law happens around $\eta\approx1$. The same kind of behavior is observed in both elections. Notice that the shape of the efficiency distribution is not specific of our electoral context, since this structure has been observed in efficiency distributions corresponding to Twitter conversations of different natures and sizes \cite{Morales20141}.
 
\begin{figure}[htbp]
  \begin{center}
  \includegraphics[width=0.47\textwidth]{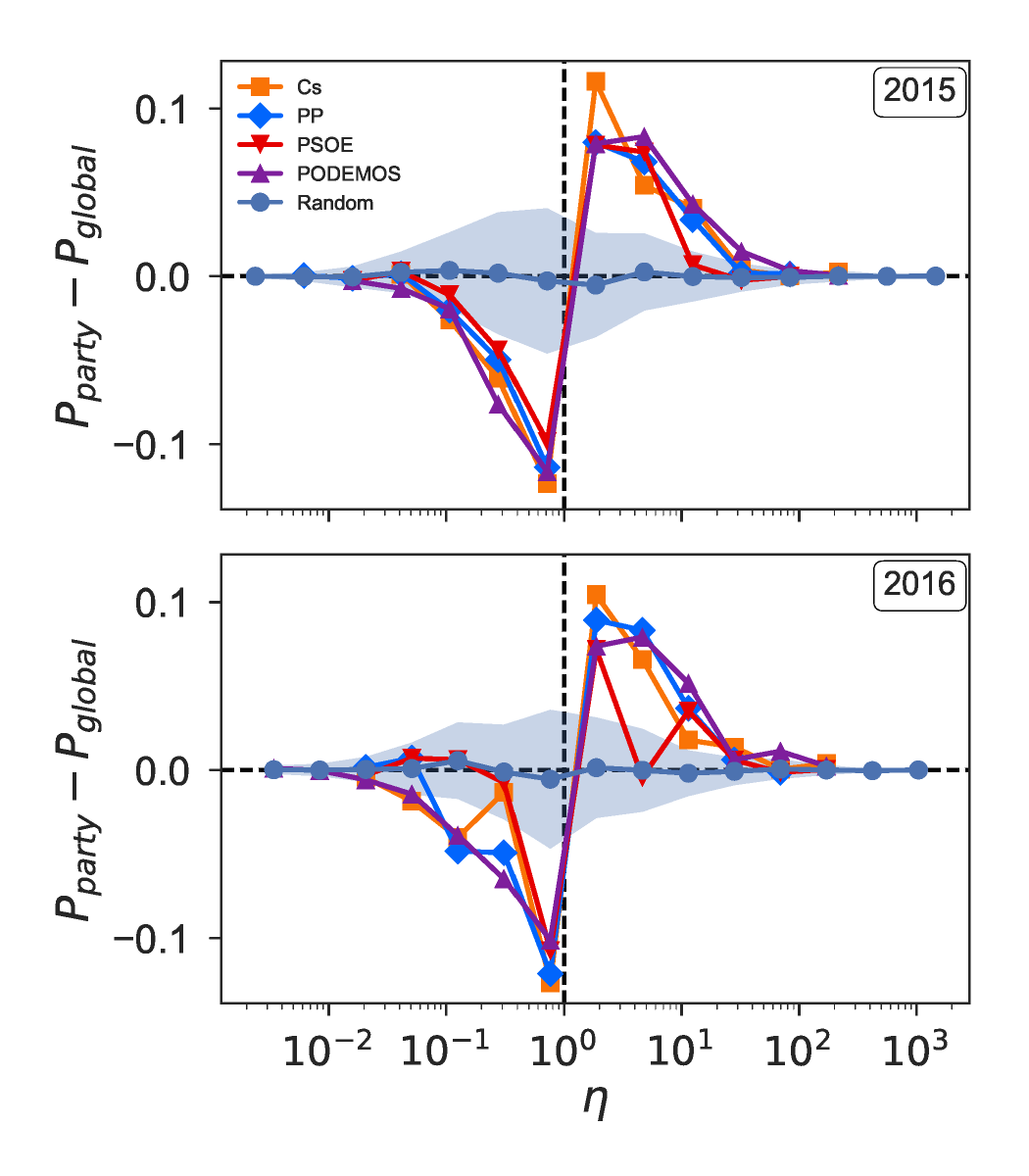}
  \end{center}	
  \caption{Probability differences of having an efficiency that falls within a given bin ($\eta_{i} < \eta < \eta_{i+1}$) between political users with respect to the total of users. The blue dots and shadowed area corresponds to the average and standard deviation of the differences for 100 samples of 1000 randomly chosen users. Political accounts tend to exhibit higher probabilities of having high efficiencies and lower probabilities of having low efficiencies.}
  \label{fig_diff_eff}
  \end{figure}

We have compared the efficiency patterns of the political accounts of each party to the whole set of users for both elections. In order to do that, we have divided the interval that spans all the possible values of efficiency (shown in figure \ref{fig_distr_eff}) in logarithmic bins. Then, we have computed the probability of finding an account of a given party in each bin ($P_{party}$) and the probability of finding a user account belonging to the global conversation ($P_{global}$). Finally, we have plotted the difference $P_{party}-P_{global}$ for the four main parties as shown in figure \ref{fig_diff_eff}. 

In order to assess the significance of this result, we have taken 100 samples of 1000 randomly chosen users and computed their efficiency probability differences with respect to the bulk of users as described above. The average values of the differences and their standard deviations are represented respectively as blue dots and a blue-grey shadow in figure \ref{fig_diff_eff}.

In that figure, it can be noticed that accounts belonging to political parties tend to exhibit higher probabilities than regular users for efficiencies in the region $\eta > 1$ and lower probabilities for efficiencies in the region $\eta < 1$. The behavior of this measure is similar for all the parties. According to this result, the collective reactions to the communication strategies of each party are comparable. Additionally, it is shown that, in general, political accounts are more efficient in propagating messages than regular users.

\subsection{Communication among the elites}

While most of the users act as passive listeners or broadcasters, Twitter conversations are usually driven by a small elite of influential accounts \cite{He2015,doi:10.1063/1.4729139}. The composition of such elite varies depending on the main subject of the conversation and on the interaction medium \cite{borondo201590}. As we have shown in the previous section, politicians are among the most efficient users, implying that they hold a high influence. Given the relevance of this group of users, in this section we will study the communication dynamics among them. To this end, we have retrieved lists of user accounts associated to each party following the process described in section \ref{sec_data}.

\begin{figure}[htbp]
  \begin{center}
  \includegraphics[width=0.5\textwidth]{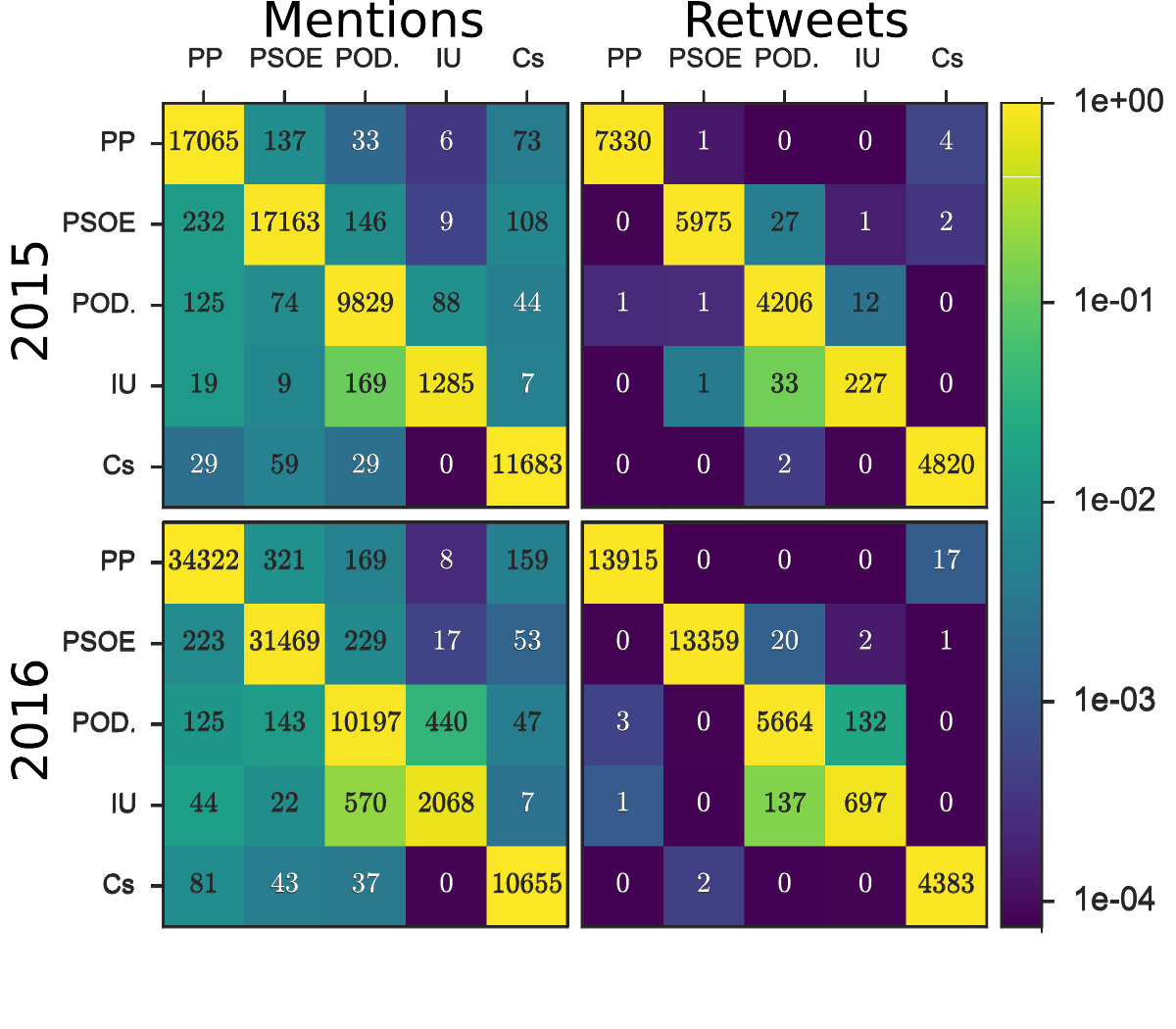}
  \end{center}	
  \caption{Adjacency matrices of C-networks where each supernode is the aggregation of political accounts belonging to a given party. Left panels correspond to mention networks and right panels to retweet networks. Top panels correspond to results of 2015 and bottom panels to results of 2016. Color is related to the proportion of mentions (retweets) directed from party $i$ (row) to party $j$ (column) by the following formula: $c_{ij} = \frac{A_{ij}}{\sum_j A_{ij}}$.}
  \label{fig_part_comm}
  \end{figure}

In order to analyze the communication among politicians, we have computed the subgraphs induced by the user accounts associated to political parties. Then, we have grouped nodes belonging to the same party in supernodes, obtaining a C-network. The resulting supernodes correspond to groups of users and the weights of the links among those supernodes are the sum of the weights of all the links that join a user from one group with a user of another in the original network. The resulting colored adjacency matrices are displayed in figure \ref{fig_part_comm}, where the numbers are the weights of the links; that is, the total number of times that users of a party in a row have mentioned (retweeted) a user of another party in a column. The color is related to the proportion of mentions (retweets) from party $i$ to party $j$ relative to the total number of mentions (retweets) by party $i$, such that if $A_{ij}$ is an element of the adjacency matrix, the color of the cell would be proportional to $c_{ij} = \frac{A_{ij}}{\sum_j A_{ij}}$.

As we can see, in line with previous works \cite{doi:10.1063/1.4729139,Yaqub2017}, there is a complete lack of communication between accounts of different parties. The diagonal of the adjacency matrix (the self-links) holds the heaviest weights by more than two orders of magnitude. For the retweet networks, there are almost no links at all between different parties. Taking into account that retweeting a message normally implies an endorsement of the ideas of the original poster, this is not surprising. In the case of the mention networks, there are more message exchanges than in the retweet network.

We have included the party IU in this analysis to study the effect of the agreement between them and Podemos to form a coalition (UP) in the 2016 election. The results displayed in figure \ref{fig_part_comm} show that, although both parties were already interacting in 2015, they exchanged much more messages in 2016.

We have also computed the evolution of the assortative mixing \cite{PhysRevE.67.026126} by political affiliation using two different partitions of the nodes. In the first partition we assign each node to its own party such that we have five groups of nodes corresponding to PP, PSOE, Cs, Podemos and IU. This is the {\it partition by party}. In the second partition, we assign the nodes belonging to IU and Podemos to the same group, leaving the rest of the nodes in their own parties. Hence, we have four groups that correspond to PP, PSOE, Cs and UP = (Podemos+IU). We have called this second partition, the {\it coalition partition}.

The assortative mixing is a metric used to test if the links of a network preferently join nodes of the same kind, nodes of different kinds or the connections are random. In our case, the different kinds would be the different parties in one case and the coalitions in the other. In order to compute the assortative mixing, the nodes are classified in groups and the proportions $e_{i j}$ of strength (that is, the total weights of the links) from group $i$ to group $j$ are computed. In our case, since we already have aggregated the nodes by party in the C-networks, we can use the elements of the  adjacency matrix $A_{ij}$ of such C-networks to determine  $e_{i j}$:

\begin{equation}
e_{i j} = \frac{A_{ij}}{\sum_{ij} A_{ij}}
 \end{equation}
 
 Then, $a_i = \sum_j e_{i j}$ is the proportion of links with an origin node of type $i$ and $b_j = \sum_i e_{i j}$ the proportion of links with a target node of type $j$. If we ignore the network structure, the probability of finding a link with origin node of type $i$ and target node of type $j$ would be $a_i \times b_j$. Taking that into account, the assortative mixing of a network is defined as follows:
 
 \begin{equation}
 r_m = \frac{\sum_i e_{i i} - \sum_i a_i b_i}{1 - \sum_i a_i b_i}
 \end{equation}
 
 This metric takes the value $r_m=1$ for a perfectly assortative network, $r_m=0$ for a random network and $r_{m}^{min}=-\frac{\sum_i a_i b_i}{1 - \sum_i a_i b_i}$ for a perfectly dissassortative one.

\begin{figure}[htbp]
  \begin{center}
  \includegraphics[width=0.5\textwidth]{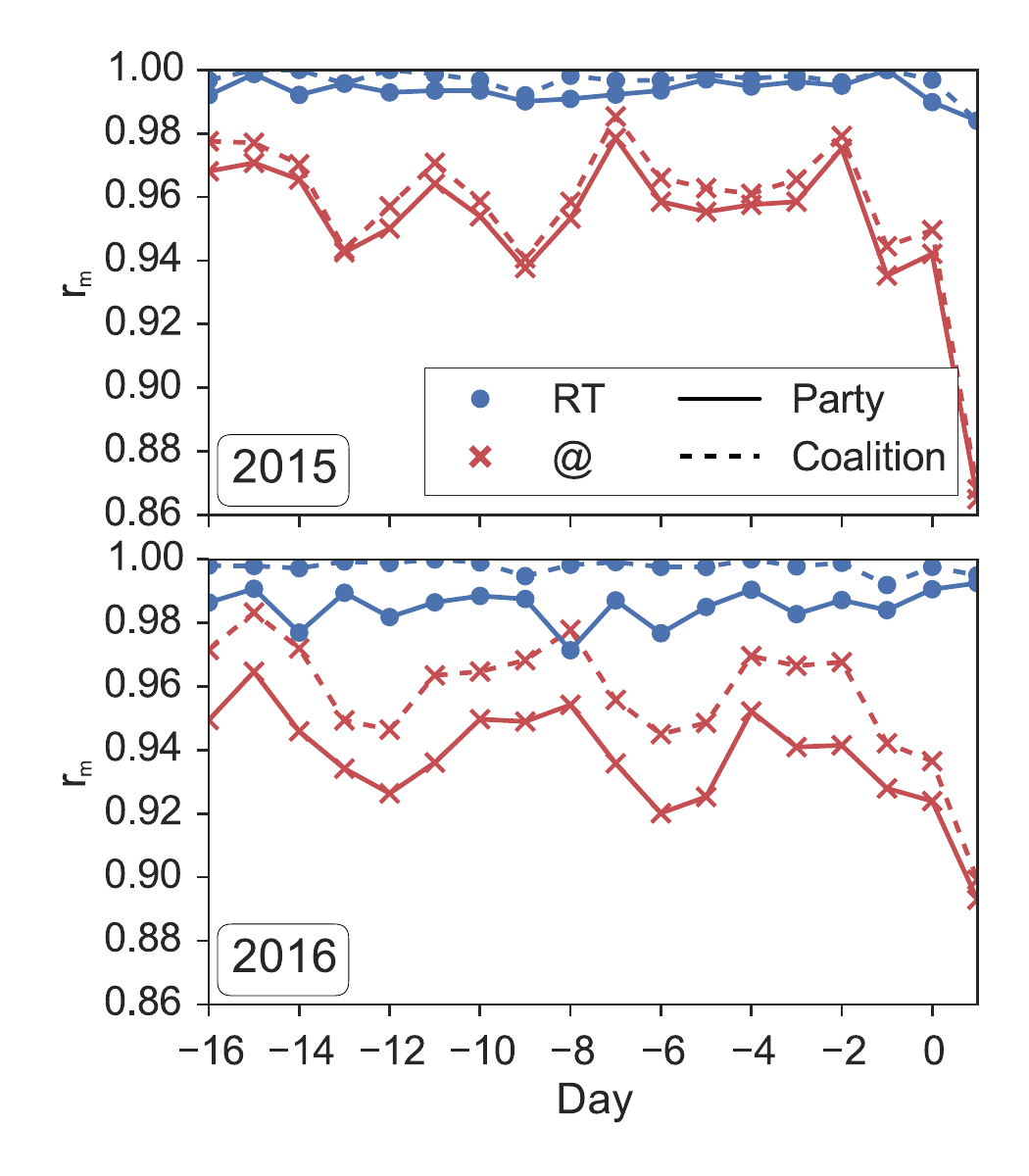}
  \end{center}
  \caption{Daily evolution of the assortative mixing by party and coalition for the mention (@) and retweet (RT) networks in 2015 (top) and 2016 (bottom) electoral campaigns.}
  \label{fig_part_assort_evol}
  \end{figure}

In figure \ref{fig_part_assort_evol} we have plotted the temporal evolution of the assortative mixing by party and coalition. As we can see, the assortative mixing by party and coalition remains very high for mentions and retweets. The retweet network is however the most assortative, often reaching values of $1.0$. 

The drop in assortativity the day after the elections means that politicians of different parties interacted more with each other that day than during the campaign. This seems to be caused partially by some exchange of messages commenting the consequences of the results of the elections. There are tweets containing criticism to adversaries, congratulation messages to related parties and tweets trying to convince or push potential allies to form coalitions. Notice, however, that this decrease, although significant, is not large: the value reached in 2015 is around 0,87 and in 2016 is 0,89. Consequently, we attribute the decrease both to the drop in the number of posted messages the day after the elections, which makes the data more noisy, and to a increment of message exchange between parties.

The most remarkable feature of these time series is the difference between the elections of 2015 and 2016. In the latter, the parties IU and Podemos formed a coalition, fact that is reflected here in the following way: whereas in 2015 coalition and party assortativities are almost equal, in 2016 the coalition assortativity is clearly higher for both networks. This means that the communication between users from Podemos and IU is high enough to lower the general assortativity. 

In order to assess the relevance of this effect, we have performed a paired t-test on the assortativities time series coupling each assortativity from 2015 with its counterpart of 2016. The null hypothesis is that the assortative mixing values for both elections have the same expected values. The results presented in table \ref{tab_ttest_assmix} imply, for a statistical confidence of $99\%$, a clear rejection of the null hypothesis for the parties time series while in the case of the coalitions time series the null hypothesis can not be rejected.

\begin{table}[htbp]
\caption{P-values of the paired t-test performed in the assortative mixing time series of parties and coalitions. The test has been carried out by taking each time series of 2015 and coupling it with its counterpart of 2016.}

\begin{tabular}{|l|c|c|c|c|}
 \hline
\multicolumn{ 1}{|l|}{\textbf{P-value}} & \textbf{Parties} & \textbf{Coalitions}  \\ \hline
Mentions & 0.0029 & 0.9  \\ \hline
Retweets & 0.0001 & 0.3  \\ \hline
\end{tabular}

\label{tab_ttest_assmix}
\end{table}

\section{Conclusions}

Our main goal is to perform a comparative analysis of the user behavior in Twitter in two consecutive electoral campaigns in order to find the presence of correlations and recurrent patterns. To this end, we have analyzed temporal series and interaction networks corresponding to two Twitter datasets downloaded during the Spanish electoral campaigns of 2015 and 2016.  Although the individual activity and the political actors may change, we have shown evidence of recurrent activity patterns in different political campaigns. In particular, the activity time series for both elections exhibit a strong correlation. Moreover, we have found a linear correlation between the daily rate of tweets and the number of unique users which is stable even when new users get into the conversation. Finally, besides the behavioral stabilities mentioned above, we have been able to detect the effect of a political coalition in the interaction networks through the study of the evolution of their properties.

The results that we have obtained from the computation and analysis of the daily user activity time series for both elections indicate that they present a significant linear correlation. Additionally, by studying the distribution of user activity we have found that in both elections its exponent fluctuates in the same tight interval. The value of the exponent obtained in a previous work \cite{doi:10.1063/1.4729139} also lies within this interval. These facts suggest the existence of recurrent activity patterns in different political campaigns.

We have shown that the daily rate of tweets, retweets and mentions follow a power law with respect to the number of unique users that participated in the conversation each day. However, whereas in 2015 the growth for the three quantities was slightly super-linear with respect to the number of users, in 2016 we observe an approximately linear behavior. Hence, in 2015, when more users join the conversation, the activity experiences a proportionally higher increment than in 2016.

We have assessed the consistency of the topology of the mentions and retweets networks from one election to the other by computing the degree distribution and the degree correlations of the aggregated networks. The variation of the power law exponent of the degree distributions from one electoral period to the other is of 1\% at most, whereas the degree correlations are shifted less than 10\% from one year to the other. The values of these properties are also comparable to the results obtained in a previous work for similar political context \cite{doi:10.1063/1.4729139}. This indicates that the underlying interaction dynamics are robust in the face of a change in social context. However, the analysis of the daily evolution of the degree assortativity of the networks has enabled us to detect  fluctuations that seem to be caused by disruptive events, both exogenous and endogenous and have a relevant impact on the political conversation.

By computing the distribution of the user efficiency for regular users and the accounts associated to each party, we have shown that its functional form is not dependent on the chosen group of users nor in the particular electoral period under study. This adds further evidence of the universality of the efficiency patterns shown by Morales et al. \cite{Morales20141}, where conversations of different natures and sizes were analyzed. We have also computed the differences between users belonging to the global conversation and politicians and found that, in the case of the latter, high efficiencies have a higher probability with respect to regular users whereas low efficiencies present lower probabilities. The behavior of this measure is similar for all the parties. According to this result, the collective reactions to the communication strategies of each party are comparable. Additionally, it is shown that, in general, political accounts are more efficient in propagating messages than regular users. Politicians are aware of the relevance of social media and know how to leverage their power.

The performed analysis of the mention and retweet C-networks induced by political accounts has enabled us to show the lack of debate among different political parties. This result is in good agreement with the existing literature \cite{doi:10.1063/1.4729139,Yaqub2017}. Furthermore, we have found that an intensification of the interaction can be detected between parties after the formation of a coalition.

In addition to the regularities in behavioral patterns that we have found by comparing two similar political contexts, several results are consistent with a previous study of the 2011 Spanish elections \cite{doi:10.1063/1.4729139}, suggesting that there exist collective behaviors that are robust in the face of social change and can be associated to the Spanish political landscape, but with a potential application beyond this social context. One possible use of these results would be to probe similar political processes and highlight anomalous behaviors that may indicate atypical Twitter uses in electoral contexts.

\section*{Acknowledgments}
This work has been supported by the Spanish Ministry of Economy and
Competitiveness (MINECO) under Contract No. MTM2015-63914-P.

An earlier version of this work has been presented at the 9th International Conference on Complex Systems.

\bibliography{bibfile}

\end{document}